\documentclass[doublecol,hyperref]{epl2}
\usepackage{epsfig,color,amsfonts}
\usepackage{xcolor}
\usepackage{ulem}
 
\title{Out-of-equilibrium fluctuations in stochastic long-range interacting systems}
\shorttitle{Out-of-equilibrium fluctuations in stochastic long-range interacting systems} 

\author{Shamik Gupta\inst{1} \and Thierry Dauxois\inst{2} \and Stefano Ruffo\inst{3}}
\shortauthor{S. Gupta \etal}

\institute{
\inst{1} Max-Planck Institute for the Physics of Complex Systems,
N\"othnitzer Stra{\ss}e 38, D-01187 Dresden, Germany \\
\inst{2} Univ. Lyon, ENS de Lyon, Univ. Claude Bernard, CNRS, Laboratoire de Physique, F-69342 Lyon, France\\
\inst{3} SISSA, INFN and ISC-CNR, Via Bonomea 265, CNISM and INFN, I-34136
Trieste, Italy}


\abstract{For a many-particle system with long-range interactions and evolving under stochastic dynamics, we study for the first time the
out-of-equilibrium fluctuations of the work done on the system by a
time-dependent external force. For equilibrium initial
conditions, the work distributions for a given protocol of variation of
the force in time and the corresponding time-reversed protocol exhibit a
remarkable scaling and a symmetry when expressed in terms of the average and the standard deviation of the work. The distributions of
the work per particle predict, by virtue of the Crooks fluctuation theorem, the
equilibrium free-energy density of the system. For a large number $N$ of
particles, the latter is in
excellent agreement with the value computed by considering the Langevin dynamics
of a single particle in a self-consistent mean field generated
by its interaction with other particles.
The agreement highlights the effective mean-field nature of the original many-particle
dynamics for large $N$. For initial
conditions in non-equilibrium stationary states (NESSs), we study the distribution of a
quantity similar to dissipated work that satisfies the non-equilibrium
generalization of the Clausius inequality, namely, the Hatano-Sasa
equality, for transitions between NESSs. Besides illustrating the
validity of the equality, we show that the distribution
has exponential tails that decay differently on the left and on the right.}
\begin{document}
\def\th{\theta}
\def\l{\label}
\def\be{\begin{equation}}
\def\ee{\end{equation}}
\def\bea{\begin{eqnarray}}
\def\eea{\end{eqnarray}}
\def\fr{\frac}
\maketitle

\section{Introduction}
Fluctuations are ubiquitous in any physical system, and
characterizing their behavior is one of the primary objectives of statistical physics.
Fluctuations may originate spontaneously or may be triggered by an
external force. When in thermal
equilibrium, the system is unable to distinguish between the two sources
of fluctuations, provided the fluctuations are small. As a result, the response of the system in thermal
equilibrium to a small external force is related to the spontaneous
fluctuations in equilibrium. The latter fact 
is encoded in the Fluctuation-Dissipation
theorem (FDT), a cornerstone of statistical physics
\cite{Kubo:1966}. Intensive
research on generalizing the FDT to situations
arbitrarily far
from equilibrium led to the formulation of a set of exact
relations, clubbed together as the Fluctuation Relations (FRs). Besides quantifying the
fluctuations, these relations constrain the entropy
production and work done on the system \cite{Seifert:2013}. Notable of the FRs are the Jarzynski equality \cite{Jarzynski:1997}
and the Crooks theorem \cite{Crooks:1999} in which the system is driven far from an initial canonical equilibrium, and the Hatano-Sasa equality \cite{Hatano:2001} that applies when the system is initially
in a non-equilibrium stationary state (NESS).

Despite such a remarkable success on the theoretical front, observing in
experiments the full range of
fluctuations captured by the FRs has been limited almost exclusively to
small systems. In a macroscopic open system comprising a large number (of the
order of Avogadro number) of constituents, the dynamics is governed
by the interaction of the environment with these many constituents, so
that any macroscopic observable such as the energy shows an average
behavior in time, and statistical excursions are but rare. A small
system, on the contrary, is one in which the energy exchange during its interaction with the
environment in a finite time is small enough so that large deviations from
the average behavior are much more amenable to observation
\cite{Bustamante:2005}. Molecular motors constitute a notable
example of small systems involved in efficiently converting chemical energy into useful mechanical work inside living cells. Recent advances in experimental manipulation
at the microscopic level led to experimentally testing the FRs, e.g., in
an RNA hairpin \cite{Collin:2005}, and in a system of microspheres optically driven through
water \cite{Trepagnier:2004}.

In this work, we consider a macroscopic system with long-range interactions that is evolving
under stochastic dynamics in presence of a time-dependent external
force. The stochasticity in the dynamics is due to the interaction of
the system with the environment. Long-range interacting (LRI) systems are those in which the inter-particle potential decays slowly
with the separation $r$ as $r^{-\alpha}$ for large $r$, with $0 \le \alpha < d$ in $d$ dimensions~\cite{review0}.

Here, we study the out-of-equilibrium fluctuations of the work done on the
system by the external force. We show that although constituted of a
large number $N$ of interacting particles, an effective single-particle
nature of the dynamics, which becomes more prominent the larger the
value of $N$ is, leads to significant statistical excursions away from
the average behavior of the work. The single-particle dynamics is
represented in terms of a Langevin dynamics of a particle evolving in a 
self-consistent mean field generated by its interaction with other
particles, and is thus evidently an effect stemming from the long-range
nature of the interaction between the particles.
For equilibrium initial
conditions, we show that the work distributions 
for a given protocol of variation of the force in time and the corresponding time-reversed protocol
exhibit a remarkable scaling and a symmetry when
expressed in terms of the average and the standard deviation of the work. 
The distributions of the work per particle predict by virtue of the Crooks theorem the
equilibrium free-energy per particle. For large $N$,
the latter value is in excellent agreement
with the analytical value obtained within the single-particle dynamics,
thereby confirming its validity.
For initial conditions in
NESSs, we study the distribution
of the quantity $Y$ appearing in the
Hatano-Sasa equality (\ref{eq:hatano-sasa}). We show that the
distribution decays exponentially with different rates on the left and
on the right.

\section{A recap of the fluctuation relations}
Consider a system evolving under stochastic dynamics, and which is characterized by a dynamical parameter $\lambda$ that can be externally controlled.
Let us envisage an experiment in which the system is subject to the
following thermodynamic transformation: starting from the stationary state corresponding to a given value
$\lambda=\lambda_1$, the system undergoes dynamical evolution under a
time-dependent $\lambda$ that changes according to a given protocol,
$\{\lambda(t)\}_{0 \le t \le
\tau};\lambda(0)\equiv\lambda_1,\lambda(\tau)\equiv\lambda_2$, over
 time $\tau$.
Only when $\lambda$ changes slowly enough over a timescale larger than the typical
relaxation timescale of the dynamics does the system pass through a succession of stationary
states. On the other hand, for an arbitrarily fast variation, the system at all times lags behind the instantaneous stationary
state. Dynamics at times $t > \tau$, when $\lambda$ does not
change anymore with time, leads the system to eventually relax to the stationary state corresponding to $\lambda_2$. In case of transitions
between equilibrium stationary states, the Clausius
inequality provides a quantitative measure of the lag at every instant of the
thermodynamic transformation between the stationary state and the actual
state of the system \cite{Bertini:2015}. For transitions between NESSs, Hatano and Sasa showed
that a quantity $Y$ similar to dissipated work measures this lag~\cite{Hatano:2001}, where $Y$ is defined as 
\be
Y\equiv\int_0^\tau \mbox{d}t~\fr{\mbox{d}\lambda(t)}{\mbox{d}t}\fr{\partial \Phi}{\partial
\lambda}(\mathcal{C}(t),\lambda(t)).
\label{eq:Y-defn}
\ee
Here, $\Phi(\mathcal{C},\lambda) \equiv -\ln \rho_{\rm
ss}(\mathcal{C};\lambda)$, and $\rho_{\rm
ss}(\mathcal{C};\lambda)$ is the stationary state measure of 
the microscopic configuration
$\mathcal{C}$
of the system at a fixed $\lambda$.
Owing to the preparation of the initial state and the stochastic
nature of the dynamics, each realization of the
experiment yields a different value of $Y$. An average over many
realizations corresponding to the same protocol $\{\lambda(t)\}$
leads to the following exact result due to Hatano and Sasa
\cite{Hatano:2001}
\be
\langle e^{-Y} \rangle=1.
\l{eq:hatano-sasa}
\ee

In the particular case in which the stationary state at a fixed $\lambda$ is
given by the Boltzmann-Gibbs canonical equilibrium state, let us
denote by $\Delta F \equiv F_2 - F_1$ the difference between the initial
value $F_1$ and the final value $F_2$ of the Helmholtz free energy that
correspond respectively to canonical equilibrium at $\lambda_1$ and $\lambda_2$. Then, if $W$ is the
work performed on the system during the thermodynamic transformation,
the Jarzynski equality states \cite{Jarzynski:1997} that 
\be
\langle e^{-\beta W}\rangle=e^{-\beta\Delta F},
\l{eq:jarzynski}
\ee
where $\beta$ is the inverse temperature of the initial canonical distribution.
Subsequent to the work of Jarzynski, a remarkable theorem due to Crooks
related (i) the distribution $P_{\rm F}(W_{\rm F})$ of the work done
$W_{\rm F}$ during the forward
process ${\rm F}$, when the system is initially equilibrated at $\lambda_1$ and inverse temperature $\beta$,  and
then the parameter $\lambda$ is changed according to the given protocol
$\{\lambda(t)\}$, to (ii) the distribution $P_{\rm R}(W_{\rm R})$ of the work
done $W_{\rm R}=-W_{\rm F}$ during the reverse process ${\rm R}$ when the system is
initially equilibrated at $\lambda_2$ and $\beta$, and
then the parameter $\lambda$ is changed according to the reverse protocol
$\{\widetilde{\lambda}(t) \equiv \lambda(\tau-t)\}$. The theorem \cite{Crooks:1999}
states that
\be
\fr{P_{\rm F}(W_{\rm F})}{P_{\rm R}(-W_{\rm F})}=e^{\beta(W_{\rm F}-\Delta
F)}.
\l{eq:crooks}
\ee
Note that the two 
distributions 
intersect at $W_{\rm F}=\Delta F$. Multiplying both sides of the
above equation by $\exp(-\beta W_{\rm F})$, and noting that $P_{\rm
R}(-W_{\rm F})$ is normalized to unity, one recovers the Jarzynski
equality.

\section{Our Model}
\l{sec:model}
Our model comprises $N$ interacting particles, labelled
$i=1,2,\ldots,N$, moving on a unit circle. Let the
angle $\th_i \in [0,2\pi)$ denote the location of the $i$-th
particle on the circle. A microscopic configuration of the system is
$\mathcal{C}\equiv\{\th_i;i=1,2,\ldots,N\}$. The particles interact through a
long-range potential
$\mathcal{V}(\mathcal{C})\equiv K/(2N)\sum_{i,j=1}^N[1-\cos(\th_i-\th_j)]$,
with $K$ being the coupling constant that we take to be unity in the
following to consider an attractive interaction \cite{note}.
An external field of strength $h$ produces a potential $\mathcal{V}_{\rm
ext}(\mathcal{C})\equiv-h\sum_{i=1}^N \cos \th_i$; thus, the total potential energy is
$V(\mathcal{C})\equiv\mathcal{V}(\mathcal{C})+\mathcal{V}_{\rm
ext}(\mathcal{C})$. The external field breaks the rotational invariance of $\mathcal{V}(\mathcal{C})$ under equal rotation applied to all the particles.

The dynamics of the system involves configurations evolving according to a stochastic
Monte Carlo (MC) dynamics. Every particle in a small time 
$\mbox{d}t \to 0$ attempts to hop to a new position on the circle. The $i$-th particle attempts with probability $p$ to move forward
(in the anticlockwise sense) by an amount $\phi;~0<\phi<2\pi$, so
that $\th_i \to \th_i'=\th_i+\phi$, while with probability $q=1-p$, it
attempts to move backward by the amount $\phi$, so that $\th_i \to
\th_i'=\th_i-\phi$. In either case, the particle takes up the attempted position with probability $g(\Delta V(\mathcal{C}))\mbox{d}t$. Here, $\Delta V(\mathcal{C})$ is the change in the potential energy due to
the attempted hop from $\th_i$ to $\th_i'$: $\Delta
V(\mathcal{C})=(1/N)\sum_{j=1}^N[-\cos(\th_i'-\th_j)+\cos(\th_i-\th_j)]-h [\cos \th_i'-\cos \th_i]$.
The dynamics does not preserve the ordering of particles on
the circle. The function $g$ has the form $g(z)=(1/2)[1-\tanh(\beta z/2)]$,
where $\beta$ is the inverse temperature. Such a form of $g(z)$ ensures
that for $p=1/2$, when the particles jump symmetrically forward and
backward, the stationary state of the system is the canonical
equilibrium state at inverse temperature $\beta$
\cite{can-note}. The case $p \ne q$
mimics the effect of an external drive on the particles to move in one preferential direction along the
circle. The field strength $h$ has the role of the
externally-controlled parameter $\lambda$ discussed in the preceding section.

The model was
introduced in Ref. \cite{Gupta:2013} as an LRI system evolving under MC dynamics. Depending on the parameters in the dynamics, the system
relaxes to either a canonical equilibrium state or a NESS. In either
case, the single-particle phase space distribution
can be solved {\it exactly} in the thermodynamic limit.

A model that has been much explored in the recent past to study static
and dynamic properties of LRI systems is the so-called
Hamiltonian mean-field (HMF) model \cite{review0}. This model
involves $N$ particles moving
on a circle, interacting via a long-range potential with the same form as $\mathcal{V}(\mathcal{C})$, and evolving
under deterministic Hamilton dynamics. The dynamics leads at long times
to an equilibrium stationary state. Our model may be looked upon as a
generalization of the microcanonical dynamics of the HMF model to a
stochastic dissipative dynamics in the overdamped regime, with an
additional external drive causing a biased motion of the particles on
the circle. The dissipation mimics the interaction of the system with an
external heat bath.

In the
Fokker-Planck limit $\phi \ll 1$, we may in the
thermodynamic limit $N \to \infty$ consider, in place of the
$N$-particle dynamics described above, the motion of a single particle in a
self-consistent mean field generated by its interaction with all the other
particles. The dynamics of the particle is described by the Langevin
equation \cite{Gupta:2013}
\be
\dot{\th}=(2p-1)\phi-\fr{\phi^2\beta}{2}\frac{\mbox{d} \langle v
\rangle}{\mbox{d} \th}+\phi \eta(t),
\l{eq:langevin-sp}
\ee
where the dot denotes differentiation with respect to time, and $\eta(t)$
is a Gaussian, white noise with $\overline{\eta(t)}=0,
~\overline{\eta(t)\eta(t')}=\delta(t-t')$. Here, overbars denote
averaging over noise realizations. In Eq. (\ref{eq:langevin-sp}), $\langle v \rangle \equiv \langle v
\rangle[\rho](\th,t)\equiv-m_x[\rho] \cos \th-m_y[\rho] \sin \th- h\cos
\th$ is the mean-field potential, with $(m_x[\rho],m_y[\rho])\equiv\int \mbox{d}\th ~(\cos \th, \sin
\th)\rho(\th,t)$, where $\rho(\th,t)$ is the probability density of the
particle to be at location $\th$ on the circle at time $t$. Together with 
$\rho(\th,t)=\rho(\th+2\pi,t)$, and the
normalization $\int_0^{2\pi} \mbox{d}\th ~\rho(\th,t)=1~~\forall~ t$,
$\rho(\th,t)$ is a solution of the Fokker-Planck equation \cite{Gupta:2013}
\be
\fr{\partial \rho}{\partial t}=-\fr{\partial}{\partial \th}
\Big[\Big((2p-1)\phi-\fr{\phi^2\beta}{2}\frac{\mbox{d}\langle v \rangle}{\mbox{d}\th}\Big)\rho\Big]+\fr{\phi^2}{2}\fr{\partial^2\rho}{\partial \th^2}.
\l{eq:FP-sp}
\ee

\section{Steady state}
Let $P(\mathcal{C},t)$ be the probability to observe configuration
$\mathcal{C}$ at time $t$. At long times, the system relaxes to a stationary state
corresponding to time-independent probabilities $P_{\rm
st}(\mathcal{C})$. For
$p=1/2$, the system has an equilibrium stationary
state in which the condition of detailed balance is satisfied, and $P_{\rm
st}(\mathcal{C})$ is given by the canonical equilibrium measure $P_{\rm
eq}(\mathcal{C}) \propto
e^{-\beta V(\mathcal{C})}$. On the other hand, for $p \ne 1/2$, the system at
long times reaches a NESS, which is characterized
by a violation of detailed balance that leads to closed loops of net non-zero
probability current in the phase space. 

For the single-particle dynamics (\ref{eq:langevin-sp}), the stationary
solution $\rho_{\rm ss}$ of Eq. (\ref{eq:FP-sp}) is given by \cite{Gupta:2013}
\be
\rho_{\rm ss}(\th;h)=\fr{\rho_{\rm
ss}(0;h)}{e^{g(\th)}}\left[1+(e^{-\fr{4\pi(2p-1)}{\phi}}-1)\fr{A(\th)}{A(2\pi)}\right]; 
\l{eq:sp-st-soln-fsame}
\ee
$g(\th) \equiv -2(2p-1)\th/\phi+\beta \langle v \rangle[\rho_{\rm
ss}](\th)$, $A(\th)\equiv \int_0^\th \mbox{d}\th'e^{g(\th')}$, while the constant $\rho_{\rm ss}(0;h)$
 is fixed by the normalization $\int_0^{2\pi} \mbox{d}\th
~\rho_{\rm ss}(\th;h)=1~\forall~h$. To show the effectiveness of the
single-particle dynamics in describing the
stationary state of the $N$-particle
dynamics for large $N$ and for $\phi \ll 1$, Fig. \ref{eq:rhoss-th-sim} shows a comparison between the result
(\ref{eq:sp-st-soln-fsame}) and MC simulation results for the $N$-particle
dynamics with $N=500,\phi = 0.1$, demonstrating an excellent
agreement.

\begin{figure}[!ht]
\centering
\includegraphics[width=5cm]{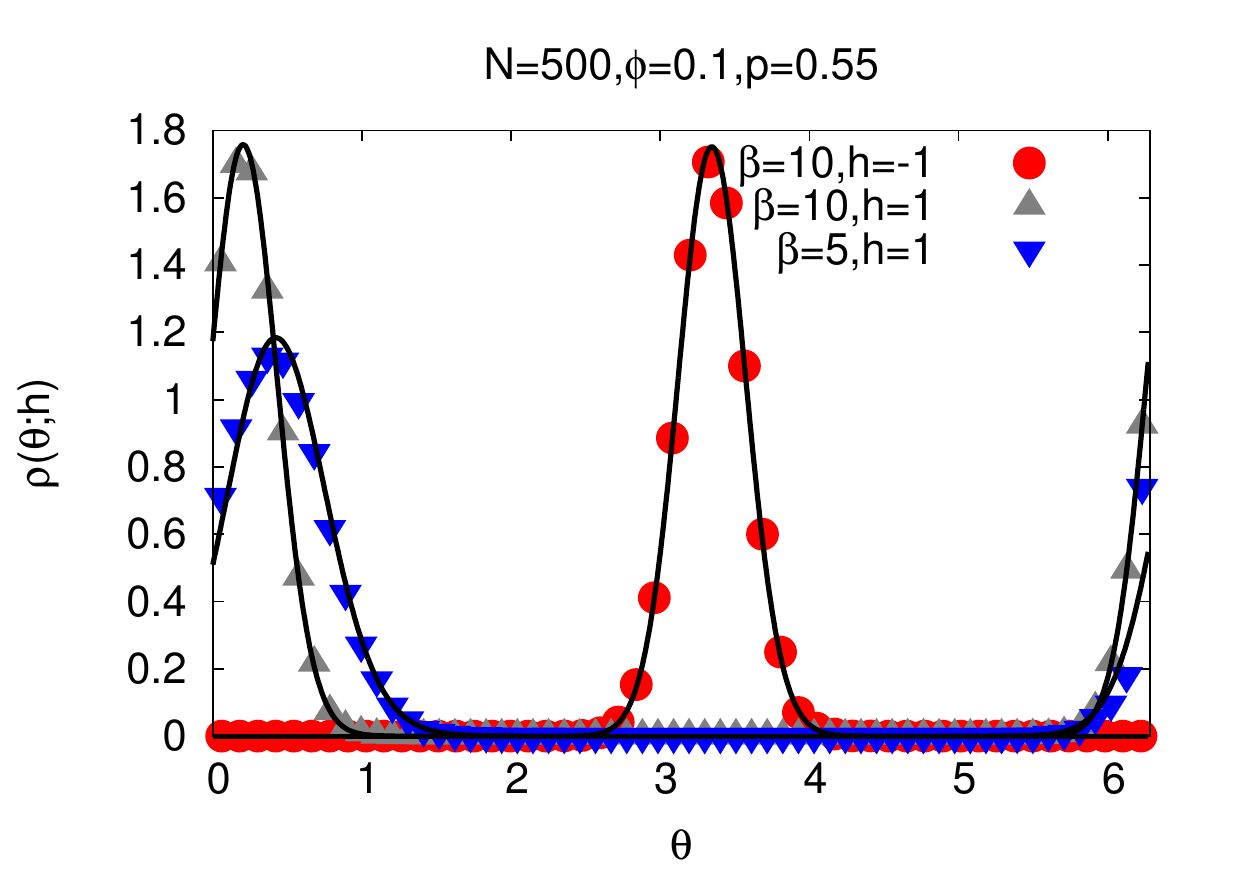}
\caption{Stationary distribution $\rho_{\rm ss}(\th;h)$ for $p \ne 1/2$: A comparison between MC simulation results
(points) for $\phi = 0.1, N=500$ and three values of field $h$, and the theoretical result
(continuous lines) in the Fokker-Planck approximation in the limit
$N \to \infty$ given by Eq. (\ref{eq:sp-st-soln-fsame}) illustrates an
excellent agreement.}
\l{eq:rhoss-th-sim}
\end{figure}

For $p=1/2$, Eq. (\ref{eq:sp-st-soln-fsame}) gives the equilibrium
single-particle distribution $\rho_{\rm eq}(\th;h)=e^{-\beta \langle v \rangle[\rho_{\rm
eq}](\th)}/Z(h)$,
with $Z(h)\equiv
\int_0^{2\pi}\mbox{d}\theta~\rho_{\rm eq}(\th;h)=2\pi
I_0(\beta m_{\rm eq})$,
and $I_n(x)$ the modified Bessel function of order $n$.
Here, $m_{\rm eq}\equiv \sqrt{(m_x^{\rm eq}+h)^2+(m_y^{\rm eq})^2}$ is obtained by solving
the transcendental equation $m_{\rm eq}+h=I_1(\beta m_{\rm
eq})/I_0(\beta m_{\rm eq})$, see~\cite{review0}. For $h=0$,
$m_{\rm eq}$ as a function of $\beta$ decreases continuously from unity
at $\beta=\infty$ to zero at the critical value $\beta_c=2$, and remains zero at
smaller $\beta$, thus showing a second-order phase transition at
$\beta_c$ \cite{review0}. For $h \ne 0$, the magnetization is
non-zero at all $\beta$, hence, there is no phase transition. 

\section{Equilibrium initial condition}
Let us consider $p=1/2$ in our model, when the system at a fixed value of $h$ has an equilibrium
stationary state. In the following,
we measure time in units of MC steps, where one MC step corresponds to
$N$ attempted hops of randomly chosen particles. Starting with the system
in equilibrium at $h=h_0$, we perform MC simulations of the
dynamics while changing the field
strength linearly over a total time $\tau \in \mathbb{I}$, with $\tau \ll
\tau_{\rm eq}$, such that at the
$\alpha$-th time step, the field value is $h_\alpha=h_0+\Delta h~\alpha/\tau;~\alpha \in [0,\tau]$. Here, $\Delta h$ is the total change in the value of the field over time $\tau$. Note that the FRs are expected to hold for arbitrary protocols $\{\lambda(t)\}$; the linear variation we consider is just a simple choice. Here,
$\tau_{\rm eq}$ is the typical equilibration time at a
fixed value of $h$, and the condition $\tau \ll
\tau_{\rm eq}$ ensures that the system during the thermodynamic transformation is driven arbitrarily far from equilibrium. The initial equilibrium configuration is prepared by
sampling independently each $\theta_i$ from the single-particle
distribution $\rho_{\rm eq}(\th;h)$, with $m_{\rm eq}$ determined by solving
$m_{\rm eq}+h_0=I_1(\beta m_{\rm
eq})/I_0(\beta m_{\rm eq})$. The 
work done on the system during the evolution is \cite{Jarzynski:1997}
\be
W_{\rm F} \equiv \int_0^\tau \fr{\partial V}{\partial h} \dot{h}\,
\mbox{d}t=-\fr{1}{\tau}\sum_{\alpha=1}^{\tau}\sum_{i=1}^N \cos \th_i^{(\alpha)},
\label{eq:Wdefn}
\ee
where $\th_i^{(\alpha)}$ is the angle of the $i$-th particle at the
$\alpha$-th time step of evolution.
In another set of experiments, we prepare the system to be initially in
equilibrium at $h=h_\tau$, and then
evolve the system while decreasing the field strength linearly in time
as $h_\alpha=h_\tau-\Delta h~\alpha/\tau$.
During these forward and reversed protocols of changing the field, we 
compute the respective work distributions $P_{\rm F}(W_{\rm F})$ and $P_{\rm
R}(W_{\rm R})$, for $\phi \ll 1$ and a number of system
sizes $N \gg 1$. We take $\tau_{\rm eq}=N^2$, confirming  
that the distributions $P_{\rm F}(W_{\rm F})$ and $P_{\rm R}(W_{\rm R})$
do not change appreciably by considering $\tau_{\rm eq}$ larger than
$N^2$.

Figures \ref{fig:crooks-inhom}(a),(b) show the forward and the
reverse work distribution for a range of system sizes $N$. Here, we have taken $\phi=0.1,h_0=1.0,\tau=10,\beta=1,\Delta h=1.0$. The data collapse evident from the plots suggests the scaling 
\be
P_{\rm B}(W_{\rm B})\sim \frac{1}{\sigma_{\rm B}}
g_{\rm B}\Big(\fr{W_{\rm B} -\langle W_{\rm B} \rangle}{\sigma_{\rm
B}}\Big);~~~~B\equiv F,R,
\l{eq:work-scaling-BHP}
\ee
where $g_{\rm B}$ is the scaling function, while $\langle W_B\rangle$ and
$\sigma_{\rm B}$ are respectively the average and the
standard deviation of the work. 
A similar scaling, termed the Bramwell-Holdsworth-Pinton (BHP) scaling, was first observed 
in the context of fluctuations of injected power in confined turbulence and magnetization fluctuations
at the critical point of a ferromagnet \cite{Bramwell:1998}. Over the years, a similar scaling has been
reported in a wide variety of different contexts, from
models of statistical physics, such as Ising and percolation models,
sandpiles, granular media in a self-organized critical state~\cite{Bramwell:2000}, to 
fluctuations in river level
\cite{Bramwell:2002}, and even in fluctuations in short
electrocardiogram episodes in humans
\cite{Bakucz:2014}. Here, the BHP scaling is shown for the first
time to also hold for work distributions out of equilibrium.
The dependence of the average and the standard deviation on the system size $N$
is shown in panels (e) and(f), respectively, with the numerically data suggesting
that $\langle W_{\rm B}\rangle \propto N$, $\sigma_{\rm F}\sim N^{a};~a \approx 0.528$,
and $\sigma_{\rm R}\sim N^b;~b \approx 0.504$.
The data collapse in (c) suggests the remarkable symmetry
\be
g_{\rm F}(x)=g_{\rm R}(-x).
\l{eq:g-scaling}
\ee
An understanding of the origin of this symmetry, and particularly,
whether it is specific to our model or holds in general, is left for future work. 
Figure \ref{fig:crooks-inhom}(d) shows the distribution of the work per
particle. Two essential features of the plots are evident, namely, (i)
significant fluctuations of the work values even for large system size, and (ii)
the forward and the reverse distribution intersecting at a
common value regardless of the system size. By virtue of the Crooks theorem (\ref{eq:crooks}), this common value should be given by the free energy difference per particle $\Delta f$ between the canonical equilibrium states of the system at field values $h_\tau$ and $h_0$. This latter quantity may be computed theoretically by knowing the free energy per particle in the limit $N \to \infty$ and at a fixed value of $h$ \cite{review0}: 
\bea
f(h)=\fr{1}{2}m_{\rm eq}^2-\fr{1}{\beta}\ln \left(\int \mbox{d}\th\, e^{\beta
[(m_x^{\rm eq}+h)\cos
\th+m_y^{\rm eq}\sin \th]}\right).
\l{eq:sp-fe}
\eea
Using the above gives $\Delta f\approx−0.725$, which is seen in Fig.
\ref{fig:crooks-inhom}(d) to match very well with the
intersection point of the forward and the reverse distribution of the work.

\begin{figure}[!ht]
\centering
\includegraphics[width=6cm]{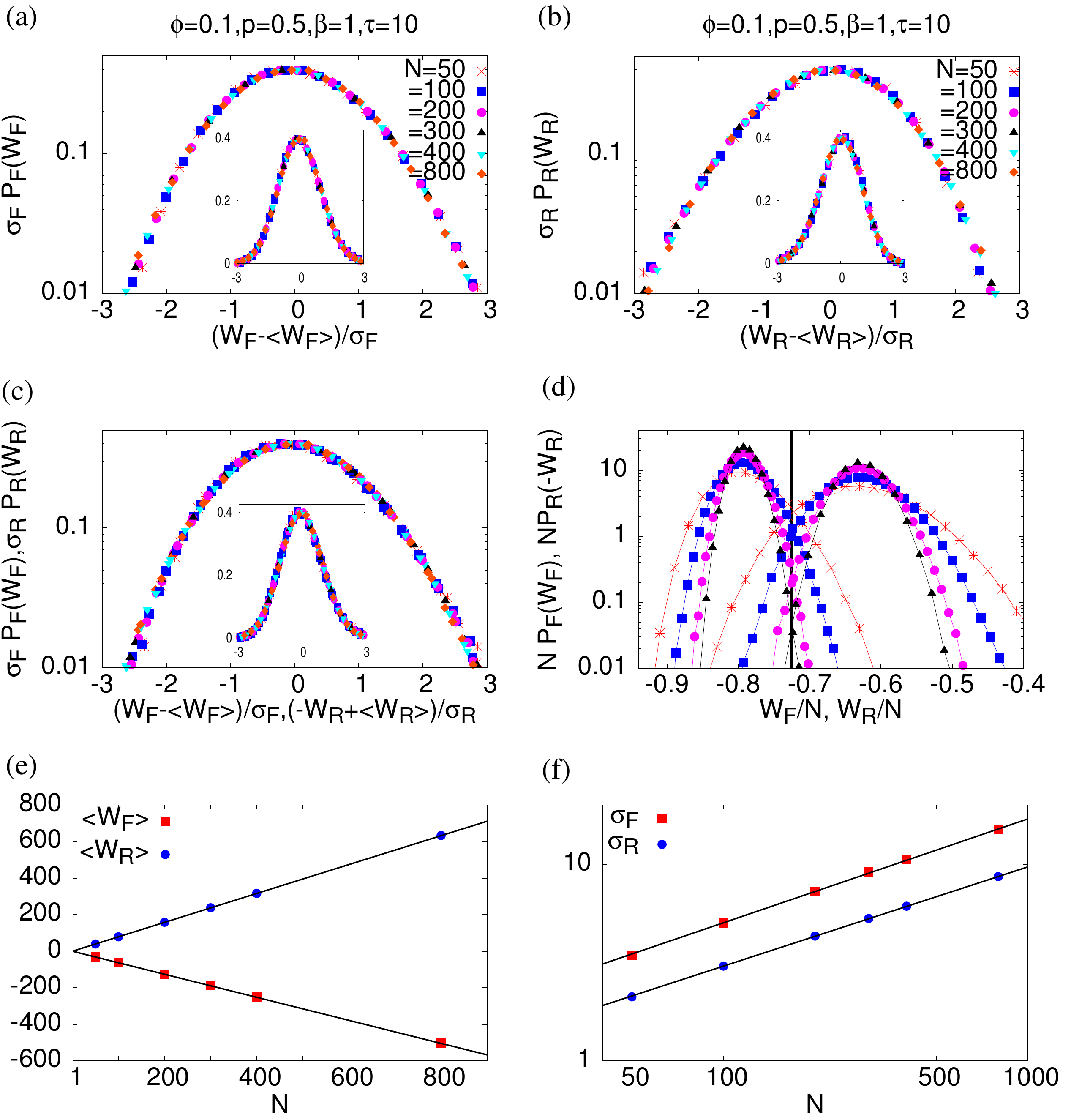}
\caption{Starting with an initial equilibrium state at inverse
temperature $\beta=1$ and field $h=h_0=1.0$, and then increasing the
field linearly in time to $h=2.0$ over a time $\tau=10$ Monte Carlo
steps (thus, $\Delta h=1.0$),
panel (a) shows the scaled work distribution for this forward
(F) protocol, while (b) shows the same for the corresponding reverse (R)
protocol, both for a range of system sizes $N$. Scaling collapse in (c)
suggests for the scaling functions in (a) and (b) the symmetry $g_{\rm F}(x)=g_{\rm R}(-x)$. (d)
shows $NP_{\rm F}(W_{\rm F})$ (right set of curves) and $NP_{\rm
R}(W_{\rm R})$ (left set) as a function
of $W_{\rm F}/N$ and $W_{\rm R}/N$, respectively, for different~$N$,
with the curves intersecting at a value given by the free
energy difference per particle $\Delta f$ estimated using Eq. (\ref{eq:sp-fe}) for
single-particle equilibrium. (e) and (f) show respectively the
dependence of the average
and the standard deviation of the forward and the reverse work on $N$,
suggesting that while the average grows linearly with $N$, one has $\sigma_{\rm F}\sim N^{a};~a \approx 0.528$,
and $\sigma_{\rm R}\sim N^b;~b \approx 0.504$. Here, $\phi=0.1,p=0.5$.}
\l{fig:crooks-inhom}
\end{figure}

\begin{figure}[!ht]
\centering
\includegraphics[width=6cm]{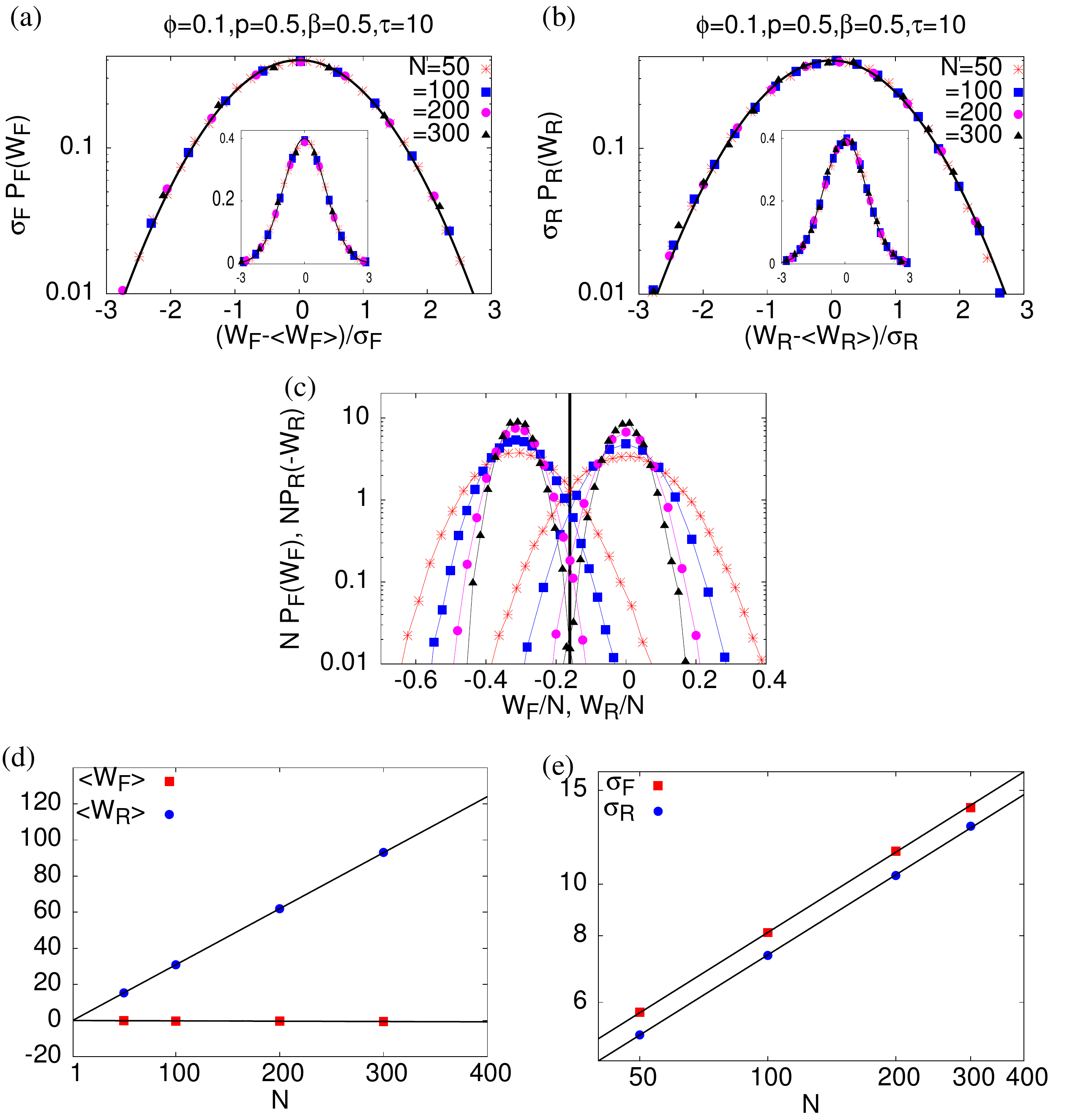}
\caption{
Plots similar to those in Fig. \ref{fig:crooks-inhom}, but with
$\beta=0.5,h_0=0.0,\tau=10,\Delta h=1.0$. The black lines in panels (a) and (b) denote a
Gaussian distribution with zero average and unit standard deviation.
While the averages in (d) grow linearly with $N$, the standard deviations in (e) satisfy $\sigma_{\rm F}\sim N^{a};~a \approx 0.501$,
and $\sigma_{\rm R}\sim N^b;~b \approx 0.5$.}
\l{fig:crooks-hom}
\end{figure}

While Fig. \ref{fig:crooks-inhom} was for inhomogeneous initial equilibrium, in order to validate our results also for homogeneous initial conditions, Figure
\ref{fig:crooks-hom} repeats the plots at
$h_0=0.0$ and at a temperature larger than $1/\beta_c$. In this case, the scaled work
distributions fit quite well to a Gaussian distribution with zero average and
unit standard deviation, see Figs.~\ref{fig:crooks-hom}(a),(b), so that
$g(x)=\exp(-x^2/2)/\sqrt{2\pi}$, and therefore, the symmetry (\ref{eq:g-scaling})
is obviously satisfied. The free energy difference $\Delta f$ can be
estimated by using Eq. (\ref{eq:sp-fe}), but can also be obtained by
using Eq. (\ref{eq:crooks}) and the fact that in the present situation, the scaled work
distributions are Gaussian. Using the latter procedure, one gets the
expression $\Delta f=(\langle W_{\rm
F}\rangle-\langle W_{\rm R}\rangle)/(2N)$ \cite{Dourache:2005}; then, on
substituting our numerical values for $\langle W_{\rm F}\rangle$ and
$\langle W_{\rm R}\rangle$, we get $\Delta f\approx −0.161$. In Fig.
\ref{fig:crooks-hom}(c), we show that this value of $\Delta f$ coincides with the point of intersection of 
the forward and the reverse work distribution.

In Figs. \ref{fig:crooks-inhom} and \ref{fig:crooks-hom}, it
may be seen that $\sigma_{\rm F} > \sigma_{\rm R}$, and also $\langle
W_{\rm R}\rangle > |\langle W_{\rm F}\rangle|$. In computing $\sigma_{\rm F}$, we start with a smaller magnetized state (thus, with the particles more spread out on the circle) than the state we start with in computing $\sigma_{\rm R}$. As a result, the work done in the former case during the thermodynamic transformation in which the increasing field tries to bring the particles closer together will show more variation from one particle to    another, resulting in $\sigma_{\rm F} > \sigma_{\rm R}$. Now, $\langle W \rangle$, either F or R, is basically the time-integrated
   magnetization, see Eq. (\ref{eq:Wdefn}). During the forward process, we start with a less-magnetized equilibrium
   state with magnetization $m_0^{\rm eq}$, and then increase the field for a finite
   time. The final magnetization value $m_{\rm fin,F}$ reached thereby
   will be smaller than the actual equilibrium value $m^{\rm eq}_1$ for the corresponding value of the field, since we did not
   allow the system to equilibrate during the transformation. For the
   reverse process, we started with this equilibrium value $m^{\rm eq}_1$, and during the
   transformation when the field is decreased, the magnetization
   decreases but not substantially to the value $m^{\rm eq}_0$, since the system remains out of
   equilibrium during the transformation. As a result, the time
   integrated forward magnetization, whose mean value is $\langle W_{\rm
   F}\rangle$, is smaller in magnitude than the time-integrated reverse
   magnetization, whose mean value is $\langle W_{\rm R} \rangle$.
   In Fig. \ref{fig:crooks-hom}, $\langle W_{\rm F} \rangle$ is very
   close to zero. This is because here, we start with a homogeneous
   equilibrium for which the magnetization value is $m^{\rm eq}_0=0$,
   and then increase the field for a finite time to a not-so-high value
   $h=1$, so the magnetization does not increase much from the initial
   value. Hence, $\langle W_{\rm F} \rangle$, which is the time-integrated magnetization during this
   forward transformation, is close to zero.

Figures \ref{fig:crooks-inhom} and \ref{fig:crooks-hom}, while illustrating
the validity of the Crooks theorem (and hence, of the Jarzynski
equality) for many-body stochastic LRI systems, underline the effective
single-particle nature of the actual $N$-particle dynamics for large $N$
in the Fokker-Planck limit $\phi \ll 1$. This feature is further
illustrated by our analysis of fluctuations while starting from NESSs,
as we now proceed to discuss.

\begin{figure}[!ht]
\centering
\includegraphics[width=4cm]{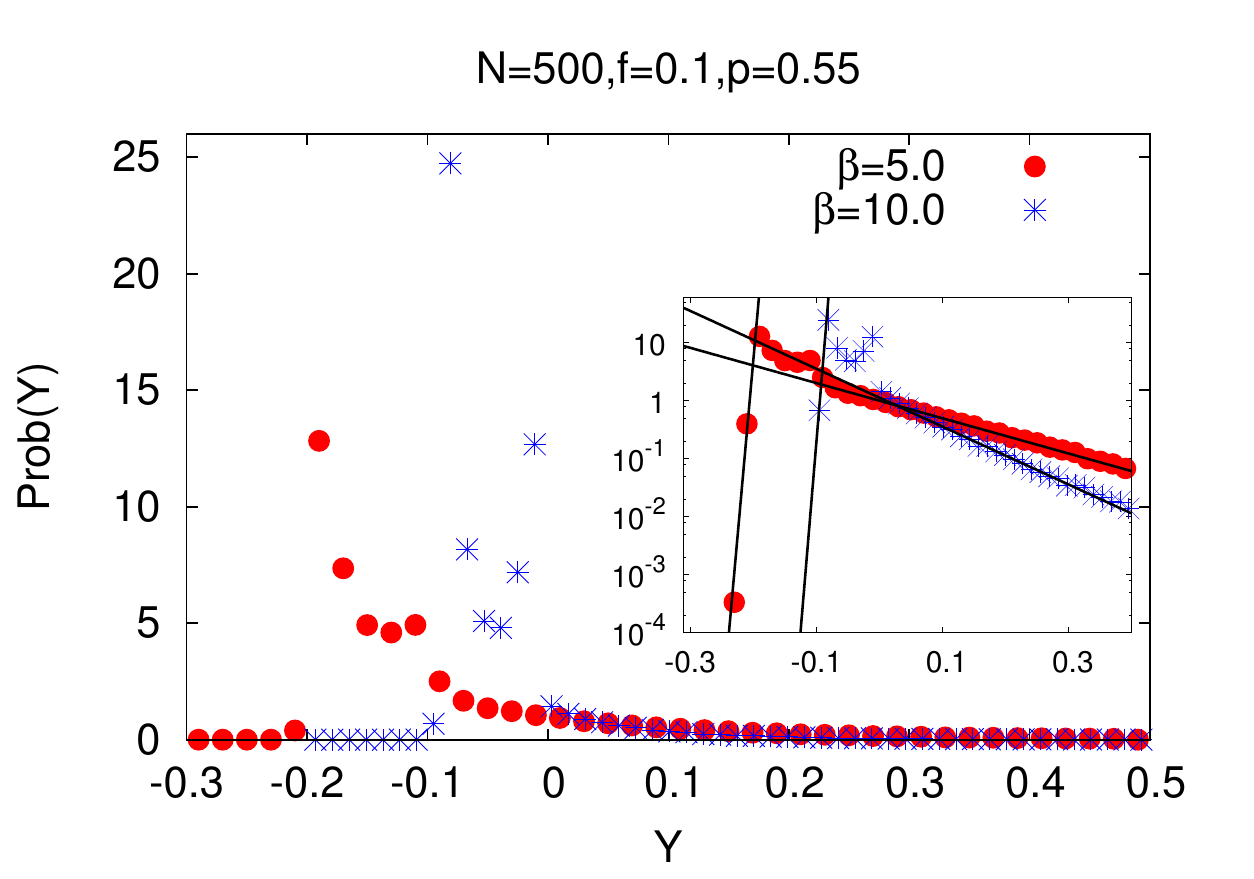}
\caption{Starting with initial conditions in a NESS at $h=h_0=1.0$, and then increasing the
field linearly in time to $h=1.15$ over $\tau=15$ Monte Carlo
steps (thus, $\Delta h=0.15$), the figure shows for two values of initial inverse 
temperature $\beta$ the distribution of the quantity $Y$ appearing in
the Hatano-Sasa equality~(\ref{eq:hatano-sasa}). The black lines in the
inset stand for the exponential fit $\sim \exp(aY)$ to the left tail,
with $a\approx 280$ for
$\beta=5$ and $a \approx 300$ for $\beta=10$, and the exponential fit
$\sim \exp(-bY)$ to the right tail, with $b\approx 7$ for
$\beta=5$ and $b \approx 11.5$ for $\beta=10$. Here,
$N=500,\phi=0.1,p=0.55$.}
\l{fig:HS}
\end{figure}

\section{Non-equilibrium initial condition}
We now consider $p\ne1/2$ in our model. In this case, the system at a
fixed value of $h$ relaxes to a NESS. We wish to compute the distribution of the quantity $Y$ appearing in the Hatano-Sasa equality (\ref{eq:hatano-sasa}). To proceed, we consider a large value of $N$ and $\phi \ll 1$ and use a combination of $N$-particle dynamics, and the
knowledge of the single-particle stationary-state distribution (\ref{eq:sp-st-soln-fsame}). Starting with the initial value $h=h_0$, the field is varied linearly in
time, as in the equilibrium case; specifically, at the
$\alpha$-th time step, the field is $h_\alpha=h_0+\Delta h~\alpha/\tau;~\alpha
\in [0,\tau]$. Again, the choice of the protocol is immaterial in as far as validity of the Hatano-Sasa equality is concerned. The steps in computing the $Y$-distribution for fixed
values of $\beta,h_0,\Delta h,\tau$ are as follows. A state
prepared by sampling independently each $\theta_i$ uniformly in
$[0,2\pi)$ is allowed to evolve under the $N$-particle MC dynamics with
$h=h_0$ to eventually relax to the stationary state, which is confirmed by checking
that the resulting single-particle distribution is given by Eq.
(\ref{eq:sp-st-soln-fsame}). Subsequently, the particles are allowed to
evolve under the time-dependent field $h_\alpha$ for a total time $\tau$, and the
quantity $Y$ is computed along the trajectory of each particle
according to Eq. (\ref{eq:Y-defn}), which is given in the present case
for the $i$-th particle by the following expression, as an approximation to the integral and the derivative appearing in Eq. (\ref{eq:Y-defn}):
\be
Y_i=\sum_{\alpha=1}^{\tau}\ln \left(\frac{\rho_{\rm
ss}(\theta_i^{(\alpha)};h_{\alpha-1})}{\rho_{\rm
ss}(\theta_i^{(\alpha)};h_{\alpha})}\right),
\label{eq:Yi}
\ee
where $\{\theta_i^{(\alpha)}\}_{0\le\alpha\le\tau}$ gives the trajectory of the $i$-th
particle, and $\rho_{\rm
ss}(\theta_i^{(\alpha)};h_{\alpha})$ is computed by using Eq. (\ref{eq:sp-st-soln-fsame}). Repeating
these steps yields the distribution of $Y$ for each particle, which is finally averaged over all the
particles to obtain the distribution $P(Y)$ depicted in Fig.
\ref{fig:HS}. Here, we use two
values of $\beta$, while the other parameters are $p=0.55,N=500,\phi=0.1,h_0=1.0,\Delta h=0.15,\tau=15$. As is evident from the
figure, the distribution is highly asymmetric, and in particular, has exponential
tails (see the inset). From the data for $P(Y)$, we find for $\langle
\exp(-Y)\rangle$ the value $1.04$ for
$\beta=10$, and the value $1.11$ for $\beta=5$, which within numerical
accuracy are consistent with the expected value of unity. Let us reiterate the combined use of the 
$N$-particle dynamics and the exact single-particle stationary state
distribution in obtaining the $Y$-distribution, and remark that the consistency of
the final results with the Hatano-Sasa equality further highlights the
effective mean-field nature of the $N$-particle dynamics for large $N$. 

To conclude, in this work, we studied the out-of-equilibrium
fluctuations of the work done by a time-dependent external force on a
many-particle system with long-range interactions and evolving under
stochastic dynamics. For both equilibrium and non-equilibrium initial
conditions, we characterized the fluctuations, and revealed how a
simpler single-particle Langevin dynamics in a mean field gives accurate
quantitative predictions for the $N$-particle dynamics for large $N$.
This in turn highlights the effective mean-field nature of the original
many-particle dynamics for large $N$.
It is interesting to generalize recent studies of work statistics in quantum
many-body short-range systems, e.g.,
\cite{Russomanno:2015,Dutta:2015}, to those with long-range interactions, and
unveil any effective mean-field description. 

SG and SR thank the ENS de Lyon for hospitality. We acknowledge
fruitful discussions with A. C. Barato, M. Baiesi, S. Ciliberto, G. Jona-Lasinio, and A. Naert.


\begin{thebibliography}{99}
\bibitem{note}The exponent $\alpha$ characterizing the decay of the inter-particle potential with separation is zero here, thus corresponding to the extreme case of long-range interactions, when the potential does not decay at all with distance.
\end{thebibliography}
\end{document}